\documentclass[conference,letterpaper,comsoc]{IEEEtran}

\usepackage[cmex10]{amsmath}
\usepackage{amsfonts, xcolor, comment, url, hyperref, cite}
\usepackage{graphicx}

\newtheorem{theorem}{Theorem}

\usepackage{tikz}

\newcommand{\note}[1]{\textcolor{blue}{[\textsf{#1}]}}

\begin{document}

\title{Pooled testing to isolate infected individuals}

%%% Single author, or several authors with same affiliation:
\author{% 
  \IEEEauthorblockN{Matthew Aldridge}
  \IEEEauthorblockA{School of Mathematics\\
                 University of Leeds\\
                 Leeds, U.K., LS2 9JT\\
                 Email: m.aldridge@leeds.ac.uk}
}

\maketitle

\newcommand{\covid}{\textsc{covid}-19}
\newcommand{\sars}{SARS-CoV-2}
\newcommand{\ETI}{\mathrm{ETI}}

\begin{abstract}
The usual problem for group testing is this: For a given number of individuals and a given prevalence, how many tests $T^*$ are required to find every infected individual? In real life, however, the problem is usually different: For a given number of individuals, a given prevalence, and a limited number of tests $T$ much smaller than $T^*$, how can these tests best be used? 

In this conference paper, we outline some recent results on this problem for two models. First, the `practical' model, which is relevant for screening for \covid\ and has tests that are highly specific but imperfectly sensitive, shows that simple algorithms can be outperformed at low prevalence and high sensitivity. Second, the `theoretical' model of very low prevalence with perfect tests gives interesting new mathematical results. 
\end{abstract}

\section{Introduction}

\subsection{The problem}

When testing individuals for a disease such as \covid, we can take a sample from each individual and test them separately; for $n$ individuals, this requires $n$ tests. Alternatively, we can pool samples together and test the pooled sample; in an ideal model, a test is positive if at least one individual in the pool is infected, and is negative otherwise. When the prevalence $p$ is low, the pooling method, known as `group testing' or `pooled testing', can identify all the infected individuals using fewer than $n$ tests, thereby making better use of scarce tests. (For background on group testing, see the recent survey paper \cite{survey}.)

Typically, the aim of group testing is to find every infected individual without erroneously declaring any uninfected individual to be infected. Thus the mathematical problem is typically this: Given the number of individuals $n$ and the prevalence $p$, how many tests $T^*$ are required to find the all the infected individuals, and what testing protocol achieves this. (There has also been some attention on the problem of finding a set that has a large overlap with the set of infected individuals \cite{survey,van} but that may not classify every single individual correctly.) However, considering applications in modern settings, especially in the current coronavirus pandemic, we propose a different goal.

In a large workplace, healthcare facility, or university, for example, there may be a very limited number of tests $T$ available each day -- certainly $T < T^*$, so there are too few tests to accurately find every infected worker. Thus the aim should be to find as many infected individuals as possible with those limited tests, so that those individuals can be removed from the workforce to isolate at home (for example).

Under this criterion, one could take samples from just a small proportion of the workforce but try to find all or almost all of the infected individuals in that subset; alternatively, one could take samples from a much larger proportion of the workforce, but be satisfied with finding a smaller fraction of the infected individuals within the subset.

In order to compare different strategies, we propose using the \emph{expected number of tests per infected individual found} (ETI) as the relevant figure of merit. That is, if a scheme uses an expected number $\mathbb ET$ of tests and finds an expected number $\mathbb EK$ of infected individuals, then the ETI is
\[ \ETI = \mathbb ET/\mathbb EK . \]
We would like this to be as large as possible.

\subsection{The models}

In this conference paper, we outline results for two different models, which we call the \emph{theoretical} and \emph{practical} models.

Both models have the following features in common:
\begin{itemize}
    \item There are a large number $n$ of individuals. This number is sufficiently large that looking at mathematical results in the limit as $n \to \infty$ is useful.
    \item The instantaneous prevalence rate $p \in (0,1)$ is known. Throughout we write $q = 1-p$. We use the \emph{i.i.d.\ prior} where each individual is infected independently with probability $p$. It will be useful to write $k = pn$ for the expected number of infected individuals. 
    \item Tests are very highly specific, in that a pool containing no infected samples is very highly likely to correctly give a negative result. We model this as the specificity being $1$; thus under our model we can be certain that a positive test contains at least one infected sample.
\end{itemize}

In our first model, the \emph{practical model}, we attempt to give a realistic model for coronavirus testing.
\begin{itemize}
  \item  The prevalence $p$ is a fixed constant as $n \to \infty$. This is because the prevalence of \covid\ is unlikely to change depending on the size of organisation being tested. For screening of \covid\ among asymptomatic people, values of the prevalence $p$ between $0.005$ and $0.1$ are likely to be of interest.
  \item Tests are only moderately sensitive, in that pools containing one or more infected samples may give an erroneous negative result. Our model for this is to say that each test containing at least one infected sample correctly gives a positive result with probability $u \in (0,1]$, independently between tests. Values of $u$ in the range $u = 0.6$ to $0.9$ are likely to be of interest for PCR tests.
  \item As many stages of testing will be impractical and slow, we limit adaptive strategies to two stages of testing.
\item There are high costs for a false positive declaration (that is, wrongly declaring a noninfected individual to be infected) --  for example, a healthcare worker may have to self-isolate for at least seven days for no reason. For this reason, false positive declarations will not be permitted, and we may only count individuals who we are certain are infected (under the above model assumptions).
  \item So that individuals can be sure they have the virus before isolating, we require suspicion that an individual is infected to be definitively confirmed with an individual non-pooled test. Thus we have a first stage of pooled testing, then a second stage of limited individual testing. This is known as `trivial two-stage testing'.
\end{itemize}

(For more detailed information on the practicality of pooled testing for \covid, with detailed consideration of real-life issues and accurate modelling, see the forthcoming book chapter \cite{book}.)

%The prevalence $p$ being fixed, independent of the number of items $n$, puts us in the so-called `linear regime'. In the linear regime, it is known that nonadaptive (that is, one-stage) testing cannot outperform individual testing \cite{aldridge-new} if we want to be certain to avoid false positive declarations, so here we concentrate on two- and later three-stage testing. (For recent results on nonadaptive testing with no false negatives and only few false positives, see \cite{heng-scarlett}.)

We give here some further justification for our consideration on trivial two-stage testing. When the sensitivity $u$ is less than $1$, it is impossible to definitively rule out individuals as definitely noninfected, since any negative tests might have been false negatives. Thus the only way to definitively confirm an individual is infected is by them receiving an individual test and that test being positive. Thus the second stage in any of our algorithms must be individual tests, as pooled tests in the final stage will be worthless under the modelling assumptions and success criteria we have set out.

In our second model, the \emph{theoretical model}, we follow the most common set-up for theoretical results on group testing -- see, for example \cite{survey, ABJ, JAS, frankfurt1, frankfurt2}.

\begin{itemize}
\item As $n \to \infty$, the prevalence $p$ scales like $p \sim n^{-(1-\alpha)}$ for $\alpha \in [0,1)$. Thus the true number of infected individuals is strongly concentrated around $k = pn \sim n^\alpha$.
\item Schemes must be fully nonadaptive, meaning all the tests are decided on in advance and are conducted in a single stage.
\item The criterion for success is that \emph{every} individual declared to be infected is indeed infected. However, we do allow failure to meet this criterion, provided that the probability of such failure tends to $0$ as $n \to \infty$.
\item The testing procedure is perfect. Any pool containing no infected individuals always gives a negative result, and any pool containing one or more infected individuals always gives a positive result.
\end{itemize}

\begin{table*}[!tbh]
\setlength{\tabcolsep}{10pt}
\renewcommand{\arraystretch}{1.8}
\caption{ETI for individual non-pooled testing with example parameters \\ of the sensitivity $u$ and prevalence $p$ in the practical model}
\label{mytab}
\begin{center}
\begin{tabular}{c|cccc}
\hline
$p$ & $u = 0.6$ & $u = 0.7$ & $u = 0.8$ & $u = 0.9$ \\
\hline
$0.1$ & $16.7$ & $14.3$ & $12.5$ & $11.1$ \\
$0.05$ & $33.3$ & $28.6$ & $25.0$ & $22.2$ \\
$0.02$ & $83.3$ & $71.4$ & $62.5$ & $55.6$ \\
$0.01$ & $167$ & $143$ & $125$ & $111$ \\
$0.005$ & $333$ & $286$ & $250$ & $222$ \\
\hline
\end{tabular}
\end{center}
\vspace{0.6cm}
\caption{ETI for pooled testing with example parameters of the sensitivity $u$ and prevalence $p$ in the practical model. \\ Values of the first-stage tests-per-individual $r$ and individuals-per-test $s$ are given in brackets as $(r,s)$.}
\label{mytab2}
\begin{center}
\begin{tabular}{c|cccc}
\hline
$p$ & $u = 0.6$ & $u = 0.7$ & $u = 0.8$ & $u = 0.9$ \\
\hline
$0.1$ & $12.4\quad (1, 5)$ & $9.93\quad (1,5)$ & $8.21\quad (1,4)$ & $6.91\quad (1,4)$ \\
$0.05$ & $18.0\quad (1,7)$ & $14.4\quad(1,6)$ & $11.8\quad (1,6)$ & $9.48\quad (2,10)$ \\
$0.02$ & $29.1\quad (1,10)$ & $23.0\quad (2,21)$ & $17.2\quad (2,19)$ & $13.3\quad (3,27)$ \\
$0.01$ & $41.7\quad (1,14)$ & $29.8\quad (2,32)$ & $22.2\quad (2, 29)$ & $16.2\quad (3, 42)$ \\
$0.005$ & $54.0\quad (2,55)$ & $38.2\quad (2,49)$ & $28.2\quad (3,76)$ & $19.7\quad (3, 68)$ \\
\hline
\end{tabular}
\end{center}
\end{table*}

\section{Results for the practical model}

We consider trivial two-stage algorithms with two parameters, $r$ and $s$, as studied in \cite{aldridge-trivial,google} for perfect noiseless tests. In the first stage, any individual that is sampled is sampled in $r$ pools, and each pool samples $s$ individuals. Typically $r$ is very small ($1$ and $2$ are the most common values, but $3$ and $4$ are sometimes used).

For individual non-pooled testing, we take $r = 1$, $s = 1$, and don't require a second stage. For all other sensible parameters we have $s > 1$, and in the second stage we retest any individuals that were positive in all $r$ pooled tests. (Although we don't consider it here, it may be worthwhile to retest individuals whose $r$ pooled tests were mostly -- but not entirely -- positive. We intend to study this in future work.)

Setting $r = 1$, $s > 1$, gives the simplest pooled algorithm, named \emph{Dorfman's algorithm}, after Robert Dorfman's original group testing procedure \cite{dorfman}. Here, the sampled individuals are split into pools of size $s$. If a pooled test is negative, it is assumed that all those individuals are noninfected. (This is not certain to be correct, but is strong evidence that retesting them is a poor use of resources compared with testing a new untested pool.) If the pooled test is positive, then those individuals are individually tested in the second stage.

For general $r, s$, the most mathematically convenient method is to fix the number of individuals $m \leq n$ to be sampled, then to choose a testing strategy uniformly at random, subject to each test having weight $s$ and each individual having weight $r$. This requires $m$ to be divisible by $r$ and $s$, but since these are typically small, this is not much of a restriction.

In practice, it can sometimes be more convenient to use a \emph{hypercube design}. We explain this first by considering the case $r = 2$, where we also use the term \emph{grid design}. Here, we take $m$ individuals where $m$ is a multiple of $s^2$. We imagine the individuals placed on square grids of size $s \times s$. Then each grid corresponds to $2s$ tests: $s$ tests each pooling the individuals in one row, and $s$ tests each pooling the individuals in one column. The grid design can be better than the random design for small $n$ (although the asymptotic performance is the same). However, $s^2$ can be quite large, so the divisibility issue can be awkward. The grid design was studied recently by Broder and Kumar \cite{google}.

For $r > 2$ the hypercube design requires $s = a^{r-1}$ for some integer $a > 1$. We place each $a^r$ individuals in an $a \times a \times \cdots \times a$ $r$-dimensional hypercube. Each test corresponds to one of the $rs$ $(r-1)$-dimensional `slices' of the hypercube. The case $r = 3$, $s = 9$, with a $27$ individuals in a $3 \times 3 \times 3$ hypercube was prominently studied in \cite{turok} -- this has nine $2$-dimensional $3\times 3$ slices: three front to back, three left to right, and three top to bottom. With a hypercube design, the divisibility restrictions are much stronger, but the extra structure can be more convenient for a laboratory to carry out.

\begin{theorem}
In the practical regime, the above algorithm has ETI
\[ \ETI = \frac{1}{pu} \]
for individual testing $r = s = 1$, and ETI
\[ \ETI = \frac{\frac rs + u^r \big(p + q (1 - q^{s - 1})^r\big)}{pu^{r+1}} \]
for $s > 1$.
\end{theorem}

\begin{IEEEproof}
We give a brief justification of this result. The result for an individual test is clear: the test is positive if the individual is infected (with probability $p$) and the test correctly gives a positive result (with probability $u$) for an ETI of $1/(pu)$.

Now consider a trivial two-stage algorithm with parameters $r$ and $s$.The number of first-stage tests-per-individual is $r/s$. 

An individual could potentially be retested in the second stage if it is either infected, with probability $p$, or it is not infected but all $r$ of its tests has one of the other $s-1$ individuals infected, with probability asymptotically $q (1 - q^{s - 1})^r$. (This would be exactly true if the test results were independent, but if an individual shares more than one pool with our given individual, that would not be the case. However, we prove in our full paper that the equation is asymptotically accurate in spite on the dependences.) If one of these criteria is fulfilled, the individual will be tested again if all $r$ tests correctly give a positive result, with probability $u^r$. Hence the expected number of second-stage tests per individual is $u^r (p + q (1 - q^{s - 1})^r)$.

An infected individual is found if it is indeed infected, with probability $p$, all $r$ pooled tests are correctly positive, with probability $u^r$, and the individual test is correctly positive, with probability $u$. All together, this is $pu^{r+1}$.

Finally, the ETI is the ratio of these two terms.
\end{IEEEproof}

Tables I and II shows the expected tests per infected individual found (ETI) for various plausible values of the sensitivity $u$ and the prevalence $p$. The numbers in brackets are optimal values of $(r,s)$, determined numerically. Note that $(1, s)$ denotes Dorfman's simple pooling algorithm.

Note that the pooled schemes are better than individual testing for all values of the parameters. The gain is biggest when the prevalence is low and the sensitivity is high. Dorfman's algorithm ($r = 1$) is best for moderately high prevalence or moderately low sensitivity.

For a similar analysis of Dorfman's algorithm that also allows for imperfect specificity, see \cite{book}.

\section{Results for the theoretical model}

Recall that in the theoretical model all tests are perfectly accurate, and the number of infected individuals is very close to $k = pn$, which scales like $k \sim n^\alpha$ with $\alpha \in [0,1)$.

Our result (stated slightly informally in this conference paper, with formalities to follow in a later paper full paper later) is the following.

\begin{theorem}
In the theoretical model, we can achieve an ETI of
\[ \ETI = \min\{\ETI_{\mathrm{full}} , \ETI_{\mathrm{saff}} \} , \]
where 
\begin{align*}
\ETI_{\mathrm{full}} &= \max \left\{  \log_2 \frac nk, \frac{1}{\ln 2} \ln k \right\} , \\
\ETI_{\mathrm{saff}} &= 2\mathrm{e} \log_2 \frac nk .
\end{align*}
\end{theorem}

\begin{IEEEproof}
We give a brief justification for this result, again with a full formal proof to appear in a later paper. We use similar techniques to those used in \cite{van} for the `find one defective item' and `approximate recovery' problems.

First, $\ETI_\mathrm{full}$ is the ETI achieved when using 
\[ T^* = \max \left\{ k \log_2 \frac nk, \frac{1}{\ln 2} k \ln k \right\} \]
tests to find all $k$ infected individuals \cite{frankfurt1,frankfurt2}. Suppose we instead have $T = cT^*$ tests, where $c < 1$. We `cut our losses' by immediately discarding all but $cn$ individuals. This subset will have very close to $ck$ infected individuals, and we can find all of them in 
\begin{align*} &\max \left\{ ck \log_2 \frac {cn}{ck}, \frac{1}{\ln 2} ck \ln ck \right\} \\
&\qquad\qquad\quad = \max \left\{ ck \log_2 \frac {n}{k}, c\frac{1}{\ln 2} k \ln k + O(k) \right\} \\
&\qquad\qquad\quad \sim c\left\{ k \log_2 \frac nk, \frac{1}{\ln 2} k \ln k \right\} , \\
&\qquad\qquad\quad = cT^* \end{align*}
tests, as required.

Second, $\ETI_\mathrm{saff}$ is achieved by an idea inspired by the SAFFRON scheme of \cite{saffron} (see also \cite{survey}). Suppose for the moment that we have $T = 2 \log_2 n/k$ tests. Take a subset of $n/k = 1/p$ items, and note that it contains exactly one infected individual with probability
\[ \frac nk p (1 - p)^{n/k - 1} = (1 - p)^{1/p - 1} \to \mathrm{e}^{-1} . \]
Number the individuals from $1$ to $n/k$, and let $\mathbf v_i \in \{0,1\}^l$ be the number $i$ written in binary, where the length is $l = \log_2 n/k = T/2$ (up to rounding). Further let $\overline{\mathbf v}_i \{0,1\}^l$ be $\mathbf v_i$ with the 0s and 1s flipped. Then individual $i$ is placed in the tests that correspond to the positions of the $1$s in the vector $(\mathbf v_i \overline{\mathbf v}_i) \in \{0,1\}^{2l} = \{0,1\}$. Note that this means each item is in exactly $l = T/2$ of the $2l = T$ tests.

If the set contains zero infected individuals, then all $T$ tests will be negative, and we can rule all the individuals to be noninfected. If exactly one of the individuals is infected, then exactly $l = T/2$ of the $T$ tests will be positive, and the first $T/2$ test outcomes will `spell out' in binary the number of the infected individual. If there are two or more infected individuals, then strictly more than $l = T/2$ of the tests are positive, and we find none of the expected items. Thus with our $T$ tests we find one infected individual if and only if the set contains exactly one infected individual, with is an expected number of $\text{e}^{-1}$. Hence the ETI is
\[ \ETI_{\mathrm{saff}} = \frac{T}{\mathrm{e}^{-1}} = \frac{2 \log_2 \frac nk}{\mathrm{e}^{-1}} = 2\mathrm{e} \log_2 \frac nk. \]

Given the actual value of $T$ we have, we split it into segments of size $2 \log_2 n/k$, and run the SAFFRON-inspired algorithm separately on each, obtaining the same ETI.

For a given parameter $\alpha$, we choose whichever out of the `full' and `SAFFRON' methods gives the best ETI. This proves the theorem.
\end{IEEEproof}

\begin{figure} 
\begin{center}
\includegraphics[width=0.5\textwidth]{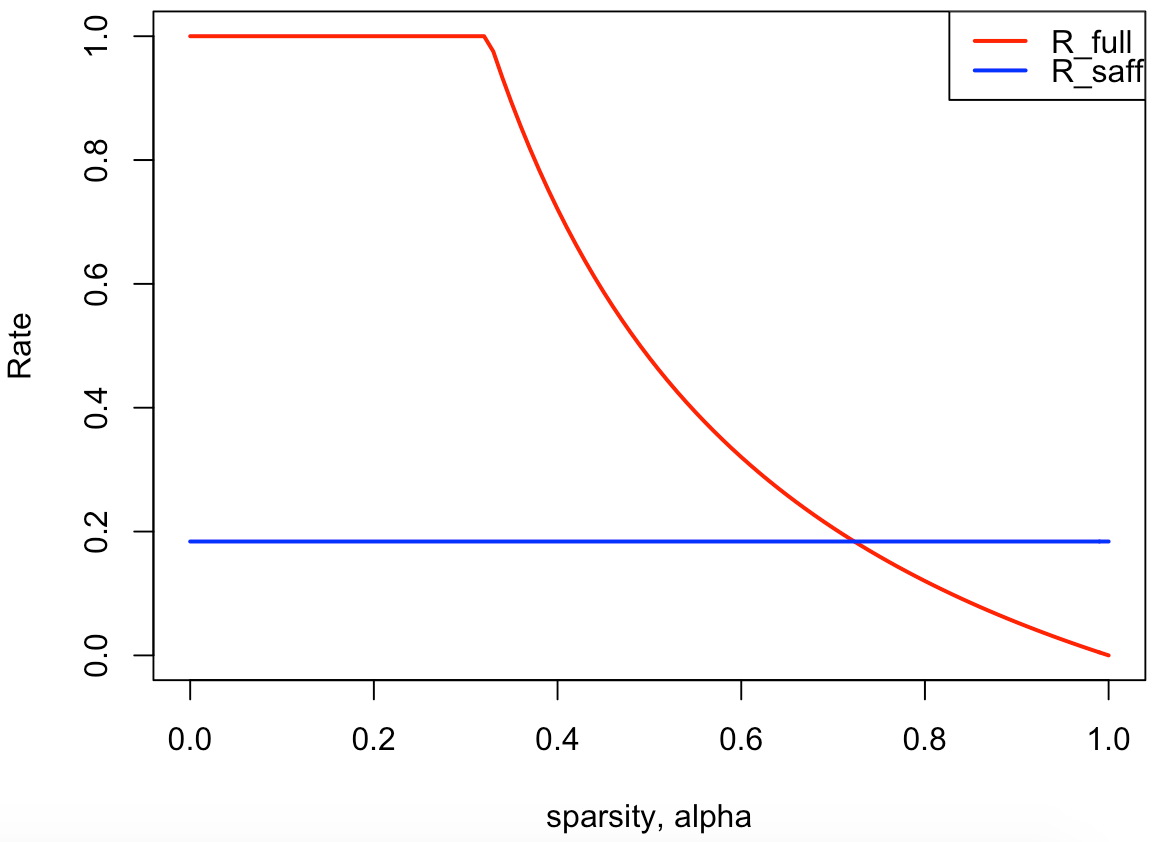}
\end{center}
\vspace{-7px}
\caption{Rate of finding isolated individuals in the theoretical model, from Theorem 3. For given $\alpha$, the better method is the one with higher rate.} \label{mainfig}
\end{figure}

Note that the `full' algorithm tests fewer individuals, but seeks to find every infected individual who was tested. In contract, the SAFFRON-inspired algorithm tests more individuals, but only expects to find $\mathrm{e}^{-1}/(1 - \mathrm{e}^{-1}) = 58\%$ of those infected.

One convenient way of interpreting Theorem 2 is by considering the \emph{rate} of group testing \cite{ABJ, survey}. Here, we define the rate to be
\[ \text{rate} = \frac{H(p)}{p\ETI} , \]
where $H(p)$ is the binary entropy. Standard information theoretic conditions tell us that the rate is bounded above by $1$; we want the rate to be as close as possible to $1$, the higher the better.

We can rewrite Theorem 2 as follows.

\begin{theorem}
When $p \sim n^{-(1-\alpha)}$, the theoretical model can achieve rates up to
\[ R = \max\{R_{\mathrm{full}} , R_{\mathrm{saff}} \} , \]
where 
\[
R_{\mathrm{full}} = \min \left\{ 1, (\ln 2)^2 \frac{1-\alpha}{\alpha} \right\} , \qquad
R_{\mathrm{saff}} = \frac{1}{2\mathrm{e}} .
\]
\end{theorem}

This is illustrated in Fig.~1. Note that, for large $\alpha$, the SAFFRON method gives a higher rate than $R_{\mathrm{full}}$, the usual rate for finding all defectives in the usual full group testing problem.

\section*{Acknowledgements}

The author thanks David Ellis for helpful comments and useful advice.

This work was supported by UK Research and Innovation under the project `Analysing group-testing algorithms for Covid-19 surveillance and case-identification in UK schools, universities and health and social care settings.'

%\section*{Acknowledgements}

\IEEEtriggeratref{8}

\bibliographystyle{IEEEtran}
\bibliography{bibliography}

\end{document}